\documentclass[structabstract]{aa}
\usepackage{savesym}
\usepackage{amsmath, bm}
\savesymbol{bibfont}
\usepackage{graphicx}
\usepackage{natbib}
\usepackage{color}
\usepackage{hyperref}
\hypersetup{
     colorlinks=true,
     linkcolor=blue,
     filecolor=blue,
     citecolor = blue,      
     urlcolor=cyan,
     }
\usepackage{txfonts}
\def\be{\begin{equation}}
\def\ee{\end{equation}}
\def\bdm{\begin{displaymath}}
\def\edm{\end{displaymath}}

\begin{document}
	   \title{Quasilinear approach of the cumulative whistler instability in 
		fast solar wind: constraints of electron temperature anisotropy}
	
		\titlerunning{Electron temperature anisotropy in the fast solar winds}
   \author{S. M.\ Shaaban \inst{1,2} \fnmsep\thanks{\email{shaaban.mohammed@kuleuven.be}} M.\ Lazar \inst{1,3}  P. H. Yoon \inst{4,5,6} \and S.\ Poedts \inst{1}
    }
         \authorrunning{S.\ M.\ Shaaban et al.}
   \institute{
	$^1$ Centre for Mathematical Plasma Astrophysics, Celestijnenlaan 200B, 3001 Leuven, 
	Belgium\\
	$^2$ Theoretical Physics Research Group, Physics Department, Faculty of Science, Mansoura University, 35516, Egypt\\
	$^3$ Institut f\"ur Theoretische Physik, Lehrstuhl IV: Weltraum- und Astrophysik, 
	Ruhr-Universit\"at Bochum, D-44780 Bochum, Germany\\
	$^4$ Institute for Physical Science and Technology, University of Maryland, College Park, MD 20742, USA\\
	$5$ Korea Astronomy and Space Science Institute, Daejeon 34055, Korea\\
	$6$ School of Space Research, Kyung Hee University, Yongin, Gyeonggi 17104, Korea
                }
   \date{Received , 2019; accepted , }
% \abstract{}{}{}{}{}
% 5 {} token are mandatory

   \abstract
   {Solar outflows are a considerable source of free energy which accumulates in multiple forms like 
	beaming (or drifting) components and/or temperature anisotropies. However, kinetic anisotropies 
	of plasma particles do not grow indefinitely and particle-particle collisions are not efficient 
	enough to explain the observed limits of these anisotropies. Instead, 
	the self-generated wave instabilities can efficiently act to constrain kinetic anisotropies, 
	but the existing approaches are simplified and do not provide satisfactory explanations. Thus,
	small deviations from isotropy shown by the electron temperature ($T$) in fast solar winds are not 
	explained yet.}
   {This paper provides an advanced quasilinear description 
	of the whistler instability driven by the anisotropic electrons in conditions typical for the 
	fast solar winds. The enhanced whistler-like
	fluctuations may constrain the upper limits of temperature anisotropy $A \equiv 
	T_\perp /T_\parallel > 1$, where $\perp, \parallel$ are defined with respect to the 
	magnetic field direction.}
   {Studied are the self-generated whistler instabilities, cumulatively driven by the temperature 
	anisotropy and the relative (counter)drift of the electron populations, e.g., core and halo electrons. 
	Recent studies have shown that quasi-stable states are not bounded by the linear instability thresholds 
	but an extended quasilinear approach is necessary to describe them in this case.}
  % results heading (mandatory)  are
   {Marginal conditions of stability are obtained from a quasilinear theory of the cumulative whistler 
	instability, and approach the quasi-stable states of electron populations reported by the observations.
	The instability saturation is determined by the relaxation of both the temperature anisotropy and 
	the relative drift of electron populations.}
   {}
	

   \maketitle
%
%________________________________________________________________

\section{Introduction}

The solar wind expansion is governed by the mechanisms which trigger the coronal plasma particles 
escape, as well as the ones self-regulating the more violent outflows, like coronal mass ejections 
or the fast winds. One of the still intriguing questions contrasting the main features of the fast
and slow winds is the electron temperature anisotropy. Deviations from isotropy shown by the 
electron temperature in the fast wind ($V_{SW} > 500$ km s$^{-1}$) are much lower than those measured 
in the slow winds \citep{Stverak2008}, suggesting the existence of an additional constraining factor. 
In the slow winds ($V_{SW} < 500$ km/s) the relative drift between thermal (core) and suprathermal 
(halo) electrons is negligibly small and it was relatively straightforward to show that large 
deviations from isotropy are mainly constrained by the self-generated temperature-anisotropy 
instabilities \citep{Stverak2008, Lazar2017a, Shaaban2019a}. In this case 
whistler modes are destabilized by a temperature anisotropy $A \equiv T_\perp /T_\parallel > 1$, where 
$\perp$ and $\parallel$ denote directions with respect to the local magnetic field.  The situation is 
however different in the fast winds, where the core-halo drift increases mainly due to an enhancement 
of the electron strahl or heat-flux population. These (counter-)drifting components can be at the
origin of the so-called heat-flux instabilities \citep{Gary1985, Saeed2017a, Shaaban2018HF, Shaaban2018HFA, 
Tong2019}, which usually are made responsible for the strahl isotropization and loss of intensity 
with solar wind expansion and increasing distance from the Sun \citep{Maksimovic2005, Vocks2005, Pagel2007}. If the 
self-generated instabilities play a role in the limitation of temperature anisotropy in this case, the 
electron strahl and heat-flux instabilities must have a contribution as well.

Heat-flux instabilities can manifest either as a whistler unstable mode, when the relative drift is small,
or as a firehose instability driven by more energetic beams with a relative drift exceeding thermal
velocity of beaming electrons \citep{Gary1985, Shaaban2018HF, Shaaban2018HFA}. Solar wind conditions 
are in general propitious to whistler heat-flux (WHF) instability \citep{Vinas2010, Bale2013, 
Shaaban2018HF, Tong2019}, and recent studies have unveiled various regimes triggered by 
the interplay with temperature anisotropy of electrons \citep{Shaaban2018HFA}. Thus, the unstable 
whistlers are inhibited by increasing the relative drift (or beaming) velocity, but are stimulated
by a temperature anisotropy $A \equiv T_\perp /T_\parallel > 1$ (with $\perp$ and $\parallel$ denoting 
directions with respect to the local magnetic field lines) \citep{Shaaban2018HFA}. 
In this paper we investigate the whistler anisotropy-driven instability for contrasting conditions in the slow 
and fast solar winds. As already mentioned, in the slow winds the relative drifts between electron 
populations are small and predictions made for non-drifting components are found satisfactory 
\citep{Stverak2008, Lazar2017a, Shaaban2019a}. On the other hand, in fast 
winds two sources of free energy may co-exist and interplay, and in this case the instability is 
cumulatively driven by temperature anisotropy under the influence of the relative core-halo drift. 
In turn, the enhanced fluctuations are expected to act and constrain the temperature 
anisotropy and eventually explain boundaries reported by the observations in the fast winds. 

However, if the instability results from the interplay of two sources of free energy threshold 
conditions provided by a linear approach may have a reduced relevance. Such a 
critical feature is suggested by recent quasilinear (QL) studies of the WHF instability driven 
by the relative drift of the core and halo electrons \citep{ Shaaban2019b}, which show the 
instability saturation emerging from a concurrent effect of the induced anisotropies, i.e., 
$A_c >1$ for the core and $A<1$ for the beam (or halo), through heating and cooling processes, 
and an additional minor relaxation of drift velocities. It becomes then clear that linear 
theory cannot describe such an energy transfer between electron populations and cannot 
characterize the quasi-stable states reached after relaxation \citep{Shaaban2018HF, Shaaban2019b}. 
Here we present the results of an extended QL analysis of the cumulative whistler 
instability, providing valuable insights about the instability saturation \textsl{via} a complex 
relaxation of the relative drift between core and halo electrons and their temperature 
anisotropy.

The present manuscript is structured as follows: The electron velocity distribution function 
(eVDF) is introduced in section~\ref{Sec:2} inspired by the observations in the slow and fast solar winds, 
which show two counter-drifting anisotropic core and halo electron populations. In section~\ref{Sec:3} 
we present in brief the QL approach used to describe a cumulative whistler anisotropy-driven 
instability in the solar wind conditions. Relevant numerical solutions are 
discussed in Section~\ref{Sec:4}, showing the influence of initial conditions (e.g., temperature 
anisotropy and (counter-)drifting velocities), on the subsequent evolutions of both the core and halo 
parameters (e.g., plasma betas, drift velocities and the associated electromagnetic fluctuations). 
Finally, the new results from our QL analysis are compared with the observational upper limits of 
electron temperature anisotropies, for both the core and halo populations and both the slow and fast 
wind conditions. The results of this work are summarized in Section~\ref{Sec:5}.
%
%%%%%%%%%%%%%%%%%%%%%%%%%%%%%%%%%%%%%%%%%%%%%%%
\section{Solar wind electrons}\label{Sec:2}
%%%%%%%%%%%%%%%%%%%%%%%%%%%%%%%%%%%%%%%%%%%%%%%
%
In the solar wind the electron velocity distribution functions (eVDF) are well reproduced by a core-halo 
model \citep{Maksimovic2005, Stverak2008, Pierrard2016, Lazar2017b, Tong2019}
\begin{equation}\label{e1}
f_e\left(v_\parallel,v_\perp \right)=\delta~f_h\left(v_\parallel,v_\perp 
\right)+ \left(1-\delta\right) ~f_c\left(v_\parallel,v_\perp \right).  
\end{equation}
where $\delta=n_h/n_0$, $1-\delta=n_c/n_0$ are the halo and core relative densities, respectively, 
and $n_0\equiv n_e$ is the total number density of electrons. A relative core-halo drift becomes 
more apparent during the fast winds ($V_{SW} > 500$ km s$^{-1}$), such that, with respect to their 
center of mass the core can be described by a drifting bi-Maxwellian, while the halo by another 
counter-drifting bi-Maxwellian or bi-kappa distribution \citep{Maksimovic2005, Stverak2008}. In 
order to keep the analysis transparent (away from suprathermal effects) here we assume both the
core (subscript $a=c$) and halo ($a = h$) populations well described by drifting bi-Maxwellians 
\citep{Saeed2017a, Tong2018, Shaaban2018HFA}  
\begin{align}
 \label{e2}
f_{a}\left( v_{\parallel },v_{\perp }\right) =&\frac{1}{\pi
^{3/2}~\alpha_{\perp~a }^{2} ~ \alpha_{\parallel~a }}\exp \left(
-\frac{v_{\perp
}^{2}}{\alpha_{\perp~a}^{2}}-\frac{\left(v_{\parallel }-U_a\right)^{2}} {\alpha_{\parallel~a}^{2}}\right),   
\end{align}
with a drifting velocity $U_a$ (along the background magnetic field) in the frame of mass-center,
e.g., fixed to protons ($a=p$), which are assumed non-drifiting Maxwellian ($U_p=0$), and $\alpha_{\parallel, \perp, 
a}(t)$ related to the components of (kinetic) temperature (which may vary in time, $t$)
\begin{align}\label{e3}
\alpha_{\perp~ a}= \sqrt{\frac{2~k_B T_{\perp~ a}}{m_e}}~~ \text{and} ~~ \alpha_{\parallel~ a}= 
\sqrt{\frac{2~k_B T_{\parallel~ a}}{m_e}}.
\end{align}
The net current is preserved zero by restricting to $n_c~U_c+~n_h~U_h=~0$ in a quasi-neutral 
electron-proton plasma with $n_e\approx n_p$. In the slow wind conditions ($V_{SW}<500$ km s$^{-1}$)  
the relative drifts are small and can be neglected, i.e.,~$U_a=0$ .

In the present analysis plasma parameterization is based on observational data provided in the last decades
by various missions, e.g., \textit{Ulysses}, \textit{Helios 1},  \textit{Cluster II} and {\it Wind} 
\citep{Maksimovic2005, Stverak2008, Pulupa2014, Pierrard2016, Tong2018}. 
In the slow wind conditions ($V_{SW}<500$ km s$^{-1}$), $U_a \simeq 0$ and the observed quasi-stable 
states and their temperature anisotropy $A = T_\perp /T_\parallel > 1$ are well constrained by the 
self-generated non-drifting whistler instability \citep{Stverak2008, Lazar2018a}. On the other hand, for the fast wind conditions ($V_{SW}>500$ km s$^{-1}$) 
the limits of temperature anisotropy $A = T_\perp /T_\parallel > 1$, for both the core and halo 
populations, are constrained to lower limits, much below the existing predictions for non-drifting 
models \citep{Stverak2008}. 

In-situ measurements also reveal a cooler electron core ($T_c<T_h$), but more dense than halo population, 
with an average number density $n_c=0.95~n_0$ \citep{Maksimovic2005, Stverak2008, Pierrard2016}. Both 
populations may display comparable temperature anisotropies, i.e., $A_c \sim A_h$, which trigger different 
instabilities such as the whistler instability ignited by $T_\perp> T_\parallel$ or firehose instabilities 
driven by $T_\parallel > T_\perp$ \citep{Pierrard2016, Lazar2018a}. Recent reports using \textit{Wind} data
suggest that the core drift velocity is comparable to, or larger than the Alfv{\'e}n speed $|U_c^w|/v_A\leqslant6$ 
(implying $|u_c^w|=\mu^{-1/2}|U_c^w|/v_A\leqslant~0.13$ at electron scales, where $\mu=~m_p/m_e$ is the proton--electron mass ratio) \citep{Pulupa2014, Tong2018}. 
As mentioned in the Introduction, the core-halo relative drift velocity can be source of beaming instabilities, 
such as, whistler heat-flux and firehose-like instabilities \citep{Gary1985, Saeed2017a, Shaaban2018HF, 
Shaaban2018HFA, Tong2019}. Values adopted here for the relative drift velocities are sufficiently 
small, e.g., $u_c=0.013\ll u_c^w $, in order to guarantee a whistler unstable regime (with a major stimulating
effect of the heat-flux, see \cite{Shaaban2018HFA}) and compare our results with the observations 
showing the limits of temperature anisotropy $A= T_\perp / T_\parallel >1$ reported in \cite{Stverak2008}.
To be consistent with our model in Eq.~\eqref{e2} we select only the events associated with thermalized 
halo components, i.e., described by a $\kappa$-distribution with large enough $\kappa > 6$.
Plasma parameters (dimensionless) adopted as initial conditions in our present analysis are tabulated in 
Table~\ref{t1}, unless otherwise specified. 
%
%   Table:1   %
\begin{table}
	\centering
	\caption{
	 Parameters for the halo and core electron populations}
   \label{t1}
	\begin{tabular}{lccc} % four columns, alignment for each
	\multicolumn{3}{c}{$\beta_{\parallel~h}(0)=0.4$, $W(k)=5\times10^{-6}$, $v_A=2\times 10^{-4}~c$}\\	
		\hline	
		 & Halo electrons ($h$)  & Core electrons ($c$)\\
		\hline
		$n_a/n_0$  & $\delta=$ 0.05 & 0.95 \\
		$T_{\parallel~a}(0)/T_{\parallel~c}(0)$ & 10.0 & 1.0 \\
		$m_p/m_a$ & 1836 & 1836\\
		$T_{\perp~a}(0)/T_{\parallel~a}(0)$ & 4.0 & 1.0, 4.0\\
		$u_a(0)$ & $u_h =$ 0.25, 0.5 & $u_c(0)=-\delta~u_h(0)/(1-\delta)$\\
		\hline
	\end{tabular}
\end{table}
%
%
%%%%%%%%%%%%%%%%%%%%%%%%%]
\section{Quasilinear Theory}\label{Sec:3}
%%%%%%%%%%%%%%%%%%%%%%%%%%
%
In a collisionless and homogeneous electron-proton plasma the linear (instantaneous) dispersion 
relation describing whistler modes read \citep{Shaaban2018HFA}
\begin{align} \label{e4}
\tilde{k}^2=&(1-\delta)~ \left[\Lambda_c+\frac{(\Lambda_c+1)~(\tilde{\omega}-\tilde{k}~u_c) - \Lambda_c}{\tilde{k} \sqrt{\beta_c}} Z_c\left(\frac{\tilde{\omega}-1-\tilde{k}~u_c}{\tilde{k} \sqrt{\beta_c}}\right)\right]\nonumber\\
&+ \delta~\left[\Lambda_h+\frac{(\Lambda_h+1)~(\tilde{\omega}-\tilde{k}~u_h) - \Lambda_h}{\tilde{k} \sqrt{\beta_h}} Z_b\left(\frac{\tilde{\omega}-1-\tilde{k}~u_h}{\tilde{k} \sqrt{\beta_h}}\right)\right]\nonumber\\
&+\frac{\tilde{\omega}}{\tilde{k}~\sqrt{\mu~\beta_p}}Z_p\left(\frac{\mu~\tilde{\omega}+1}{\tilde{k}\sqrt{\mu~\beta_p}}\right),
\end{align}
where $\tilde{k}=kc/\omega_{p,e}$ is the normalized wave-number $k$, 
$c$ is the speed of light, $\omega_{p, e}=\sqrt{4\pi n_0 e^2/m_e}$ is the plasma 
frequency of electrons, $\tilde{\omega}=\omega/|\Omega_e|$ is the normalized wave frequency, $\Omega_e$ is the non-relativistic gyro-frequency of electrons, $\Lambda_a=~A_a-1$, $\beta_{\parallel,\perp , a}=~8\pi n_0 k_B T_{\parallel \perp , a}/B_0^2$ are the plasma beta parameters for protons (subscript "$a=p$"), electron core (subscript 
"$a=c$"), and electron halo (subscript "$a=h$") populations, 
$u_a=\mu^{-1/2}~U_a/v_A$ are normalized drifting velocities, $v_A=\sqrt{B_0^2/4\pi n_p m_p}$ is the Alfv{\'e}n speed, and
\begin{equation}  \label{e5}
Z_{a}\left( \xi _{a}^{\pm }\right) =\frac{1}{\sqrt{\pi}}\int_{-\infty
}^{\infty }\frac{\exp \left( -x^{2}\right) }{x-\xi _{a}^{\pm }}dt,\
\ \Im \left( \xi _{a}^{\pm }\right) >0, 
\end{equation}
is the plasma dispersion function \citep{Fried1961}. 

Beyond the linear theory, we solve quasilinear (QL) equations for both particles and electromagnetic 
waves. For transverse modes propagating parallel to the magnetic field, the particle kinetic 
equation for electrons in the diffusion approximation is given by \citep{Seough2012, Shaaban2019b}
\begin{align} \label{e6}
\frac{\partial f_a}{\partial t}&=\frac{i e^2}{4m_a^2 c^2}\int_{-\infty}^{\infty} 
\frac{dk}{k}\left[ \left(\omega^\ast-k v_\parallel\right)\frac{\partial}{v_\perp~\partial v_\perp}+ 
k\frac{\partial}{\partial v_\parallel}\right]\nonumber\\
&\times~\frac{ v_\perp^2~\delta B^2(k, \omega)}{\omega-kv_\parallel-\Omega_a}\left[ 
\left(\omega-k v_\parallel\right)\frac{\partial}{v_\perp~\partial v_\perp}+ k
\frac{\partial}{\partial v_\parallel}\right]f_a
\end{align}
where $\delta B^2(k)$ is the energy density of the fluctuations described by the wave kinetic equation 
\begin{equation} \label{e7}
\frac{\partial~\delta B^2(k)}{\partial t}=2 \gamma_k \delta B^2(k),
\end{equation}
and $\gamma_k$ is the instability growth rate calculated from Eq.~\eqref{e4}. 
Dynamical kinetic equations for the velocity moments of the distribution function, such as temperature 
components $T_{\perp, \parallel~a}$ of core ($a=c$) and halo ($a=h$) and their drift velocities $U_a$, are 
given by
\begin{subequations}\label{e8}
\begin{align}
\frac{dT_{\perp a}}{dt}&=\frac{\partial}{\partial t}\int d{\bf{v}} ~m_a v_{\perp}^2~f_a/2\\
\frac{dT_{\parallel~a}}{dt}&=\frac{\partial}{\partial t}\int d{\bf{v}}~m_a (v_{\parallel}-U_a)^2~f_a\\
\frac{d U_a}{d t}&=\frac{\partial}{\partial t}\int d{\bf{v}}~ v_{\parallel}~f_a
\end{align}
\end{subequations}
Detailed derivations of Eqs.(\ref{e8}) can be found in \cite{Seough2012, Yoon2012, Sarfraz2016, Lazar2018b, 
Shaaban2019a, Shaaban2019b}. 

   \begin{figure*}
   \centering
\includegraphics[scale=1., trim={2.5cm 10.8cm 2.cm 2.6cm}, clip]{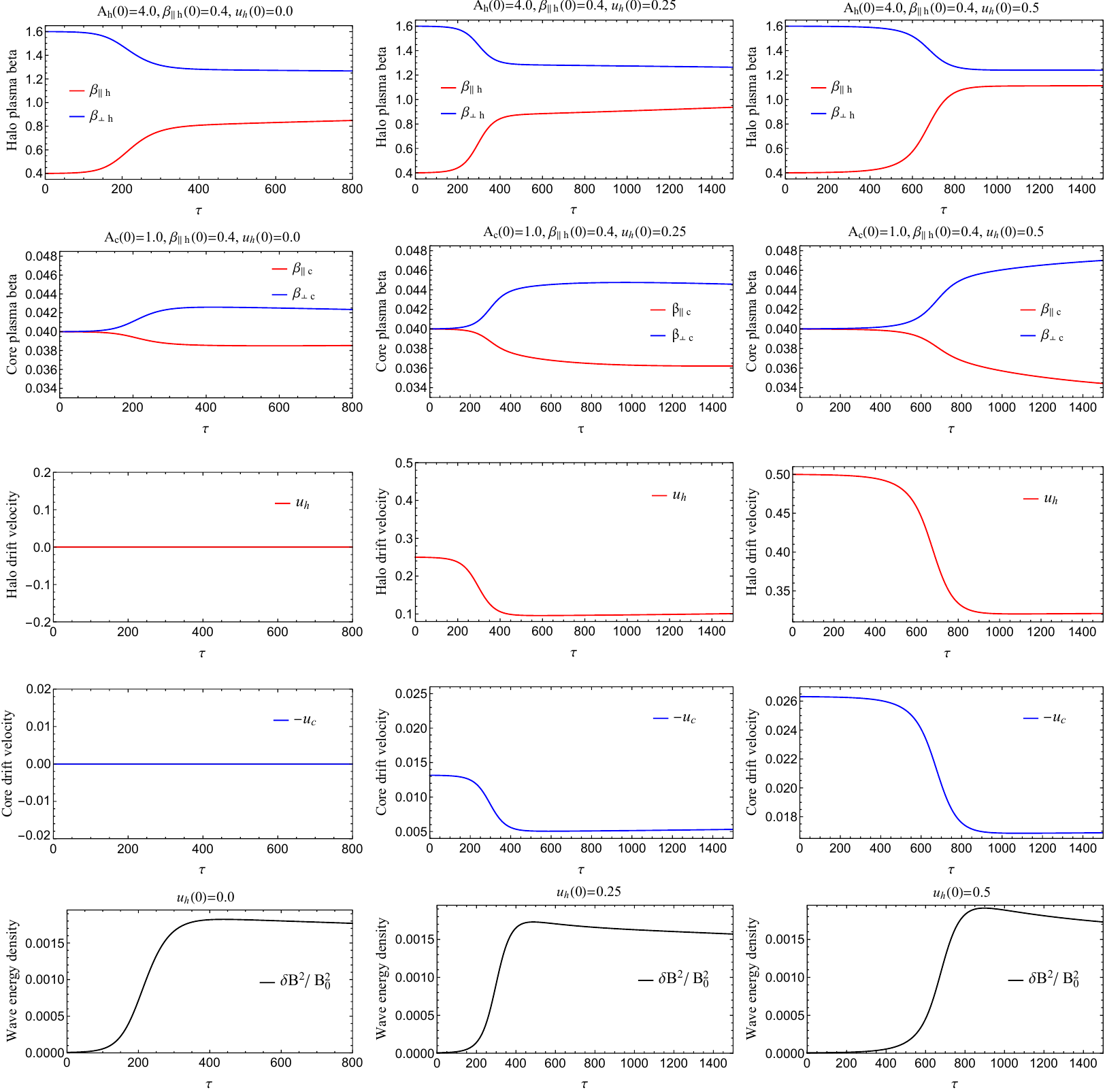}
\caption{Time evolutions for three distinct initial conditions $u_h(0)=0.0$ (left column), $u_h(0)=0.25$ 
(middle column), and $u_h(0)=0.5$ (right column), 
for $\beta_{\perp~a}$ (solid blue) and , and $\beta_{\parallel~a}$ (solid red) for core ($a=c$) 
and halo electrons ($a=h$), and their drift velocities $u_a$ (dashed), and the wave energy density 
(solid black).} \label{Fig1}
   \end{figure*}
%
%
%%%%%%%%%%%%%%%%%%%%%%%%%%%%%%%
\section{Cumulative whistler instability}\label{Sec:4}
%%%%%%%%%%%%%%%%%%%%%%%%%%%%%%
In this section we examine the whistler instability (WI) cumulatively driven by 
anisotropic temperature ($A_a>1$) and (counter-)drifting motion of the core and halo 
electrons. For electrons with anisotropic temperature $A_e > 1$ and no 
drifting components, both theory and simulations predict the excitation of parallel 
whistlers that rapidly consumes most of the free energy before the electron mirror 
instability can grow with a considerable growth rate \citep{Gary2006, Ahmadi2016}. In turn, the electron anisotropy is bounded by whistler instability
that scatters electrons and imposes a $\beta_\parallel$-dependent upper bound on the 
instability thresholds \citep{Gary2006}, which shapes the anisotropy limits 
reported by the observations in the slow wind \citep{Stverak2008, Adrian2016}. 
Motivated by these premises, and by the assumption of small drifts, which inhibit mirror modes but may stimulate whistlers \citep{Shaaban2018HFA}, in the present study we analyze only the 
whistler modes propagating parallel to the magnetic field. Remember that a whistler heat-flux 
instability (WHFI) can be excited only for a modest drift velocity, lower than thermal speed, i.e.\ $u_h<v_{res}$ 
\citep{Gary1985, Shaaban2018HF}. Both WI and WHFI have the same dispersive characteristics, being triggered by the 
resonant halo electrons and exhibiting maximum growth rates for parallel propagation \citep{Gary2006, Shaaban2018HFA}. 
Linear properties of these two distinct regimes have recently been contrasted and discussed in detail in 
\cite{Shaaban2018HFA}, showing that all dispersive features are markedly altered by the interplay of
electron anisotropies, namely their temperature anisotropy and drifting velocities. In this section we present 
the results from an extended QL analysis able to characterize the long-run time evolution of the enhanced 
fluctuations, including their saturation and back reaction on both the core and halo populations.

%%%%%%%%%%%%%%%%%%%%%%%%%%%%%%%%%%%%%%%%%%%%%%%%%%%%
\subsection{Quasilinear results} \label{Sec:4.1}
%%%%%%%%%%%%%%%%%%%%%%%%%%%%%%%%%%%%%%%%%%%%%%%%%%%%
We resolve the set of QL equations (\ref{e7}) and (\ref{e8}) for the initial parameters ($\tau=0$) in 
Table~\ref{t1}.
To do so, we use a discrete grid in the positive normalized wave-number 
$\tilde{k}-$space, i.e.\  $0.1~<~ \tilde{k} <~1.6$, with $N_k=400$ points separated by 
$d\tilde{k} = 0.04$. The initial wave spectrum is assumed to be constant over the initially 
unstable $\tilde{k}-$space \citep{Yoon1992}, and here we have adopted an arbitrary value of 
$5\times 10^{-6}$. The linear (instantaneous) dispersion relation (\ref{e4}) for WI can be solved 
using the plasma parameters and the magnetic wave energy at time step $\tau=t~|\Omega_e|$. 
The results at this moment in time are the unstable solutions of the WI, i.e.,  growth rate 
and wave-frequency as functions of $\tilde{k}$. By using these solutions we then compute
the integrals defined in the QL equations and evaluate the time derivative of each plasma 
parameters for both the core and halo populations. Then we allow the whole system to evolve 
to the next step $\tau+d\tau$ by using the second order leapfrog-like method. Numerical 
integration is performed between $0\leqslant\tau\leqslant1500$ with time step of $d\tau=0.1$. 
It is well known that such plasma systems conserve momentum and energy (that corresponds 
to the sum of the plasma particles energies, i.e., temperatures and drifts velocities, and 
the plasma waves energy). Excess of plasma particle free energies drives instabilities  
enhancing the electromagnetic fluctuations, i.e., the wave energy. In turn, the enhanced 
fluctuations scatter particles to the quasi-stable state, reducing the anisotropy 
of particle distributions through wave-particle interaction 
\citep{Moya2011, Yoon2012, Shaaban2019b, Lopez2019}.
Figure~\ref{Fig1} displays temporal profiles for the plasma beta parameters ($\beta_{\perp, \parallel}\equiv 
8\pi n_0 k_B T_{\perp, \parallel}/B_{0}^2$), parallel (red) and perpendicular (blue) components for the halo 
(subscript $h$) and core (subscript $c$) populations, and their drift velocities $u_{h,c}$, as well as the 
corresponding increase of the wave magnetic power $\delta B^2/ B_0^2$ (bottom). Here we consider the 
core initially isotropic ($A_c(0)=1$), and halo with $A_h(0)=4.0$ and initial drifts defining three cases:
$u_h(0)=~0.0$ (left), 0.25 (middle), and 0.5 (right). The effects of whistler fluctuations on the core 
and halo populations can be explained by the resonant heating and cooling mechanisms, combined with an 
adiabatic scattering and diffusion of particles in velocity space. Initially anisotropic, the halo electrons 
are subject to perpendicular cooling (blue) and parallel heating (red), while the isotropic core electrons 
experience perpendicular heating (blue) and parallel cooling (red). After relaxation, both components end 
up with small (and similar) temperature anisotropies. Moreover,  drift velocities reduce in time to lower 
but finite values. It is worth to outline the advantage of using a QL analysis that can unveil the energy 
transfer between electron populations during the instability excitation, e.g., the initially isotropic core 
gains free energy, i.e., temperature anisotropy in perpendicular direction $\beta_{\perp~c}> 
\beta_{\parallel~c}$. If the halo drift velocity is initially higher, i.e.\ $u_h(0)=0.5$ (right panels), all 
temporal variations change, namely, the halo becomes less anisotropic, the initially isotropic core ($A_c(0)=
1.0$) becomes more anisotropic ($A_c>1.0$), and the level of saturated fluctuations increases. By comparison to the first case without drifts ($u_b=0.0$),  
all these processes (and mechanisms involved) are delayed in time by a factor of $\sim1.5$ for 
$u_b=0.25$, and $\sim3$ for $u_b=0.5$.  This delay is consistent with predictions from 
linear theory that shows an inhibition of whistlers, with maximum (peaking) growth rates 
decreasing for finite drifts, by a factor of $\sim1.5$ for $u_b=0.25$, and $\sim3$ for $u_b=0.5$, 
see, for instance, figure 3 in \cite{Shaaban2018HFA}.
   \begin{figure*}
   \centering
\includegraphics[scale=1, trim={2.5cm 11cm 2.cm 2.6cm}, clip]{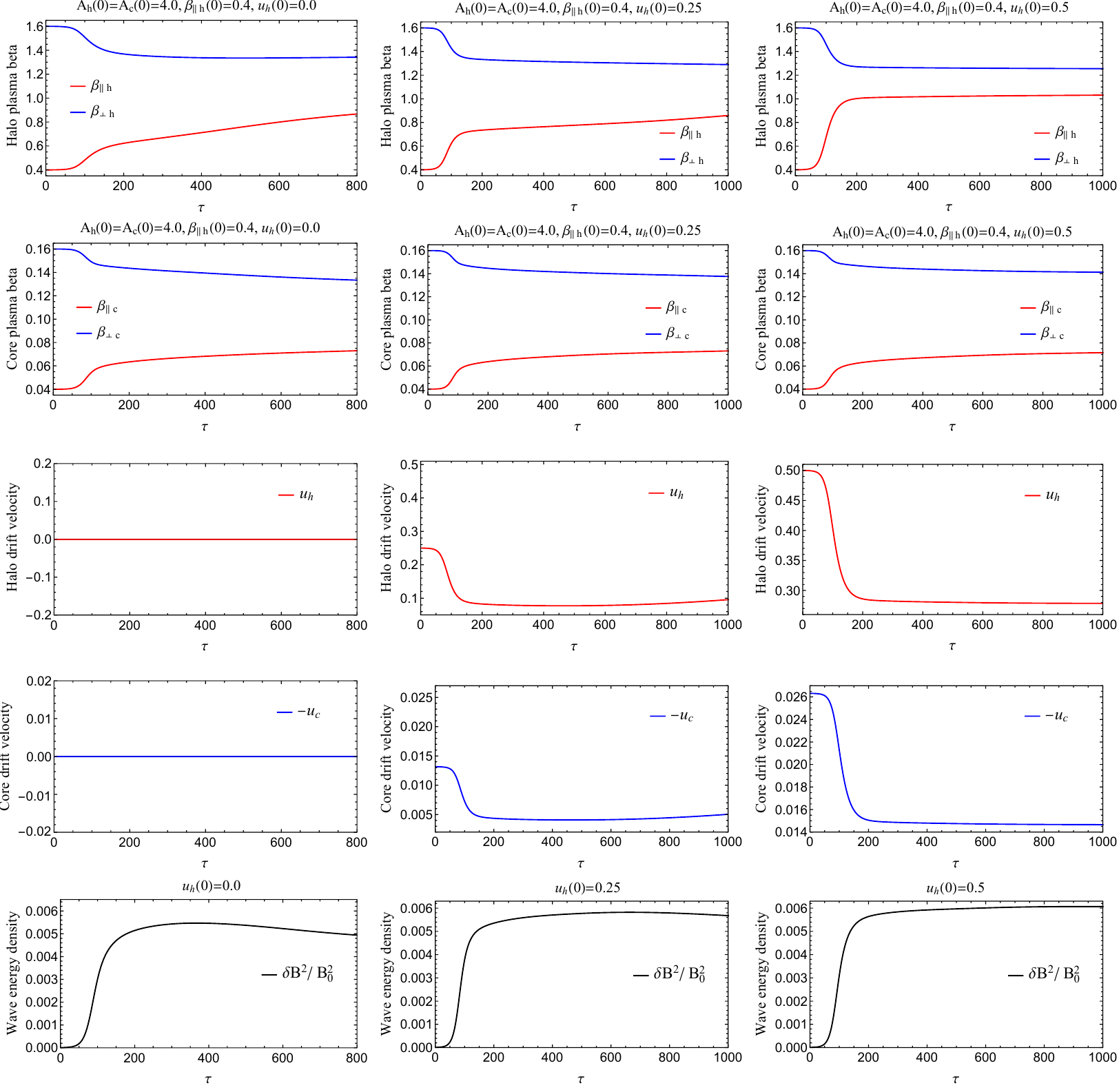}
\caption{The same as in Figure~\ref{Fig1} but for an initially anisotropic core $A_c(0) \simeq A_h(0)$.} \label{Fig2}
   \end{figure*}

Figures~\ref{Fig2} presents temporal profiles for the same plasma parameters, but for different initial 
conditions assuming that both halo and core electrons possess similar initial anisotropies $A_h (0) \simeq 
A_c (0)=4.0$. The rest of the initial plasma parameters are the same as in Figure~\ref{Fig1}. 
In this second case, both the core and halo electrons reduce their temperature anisotropies and 
drift velocities, leading to higher wave powers at the saturation. An initially anisotropic core 
($A_c (0) =4.0$) determines a faster relaxation. These results are in perfect agreement 
with recent studies which strongly suggest that electron kinetic instabilities are markedly stimulated 
by the interplay of the core and halo temperature anisotropies \citep{Lazar2018a, Shaaban2019a}. 
A higher initial drift $u_h(0)$=0.5 has visible consequences on temperature anisotropy of the halo 
electrons which decreases to lower values, but only slightly affects the time evolution of the 
core anisotropy. Predictions from linear theory show a similar influence of the drift velocity on the 
WI driven by the temperature anisotropy: Increasing the halo drift velocity $u_h$ stimulates the 
instability driven by the core anisotropy ($A_c>1$), but inhibits the fluctuations triggered by 
the halo anisotropy ($A_h>1$) \citep{Shaaban2018HFA}. For cases with $u_h\neq0$, the relaxation of the drift velocity contributes
to isotropization of the halo electrons and to an increase of the core anisotropy that feeds 
the instability and enhances the wave energy. This complex scenario may be explained by a
transfer of energy and momentum between core and halo electrons, and the enhanced fluctuations.

  \begin{figure*}[t]
   \centering
\includegraphics[scale=0.75, trim={4.2cm 6.3cm 2.6cm 6.8cm}, clip]{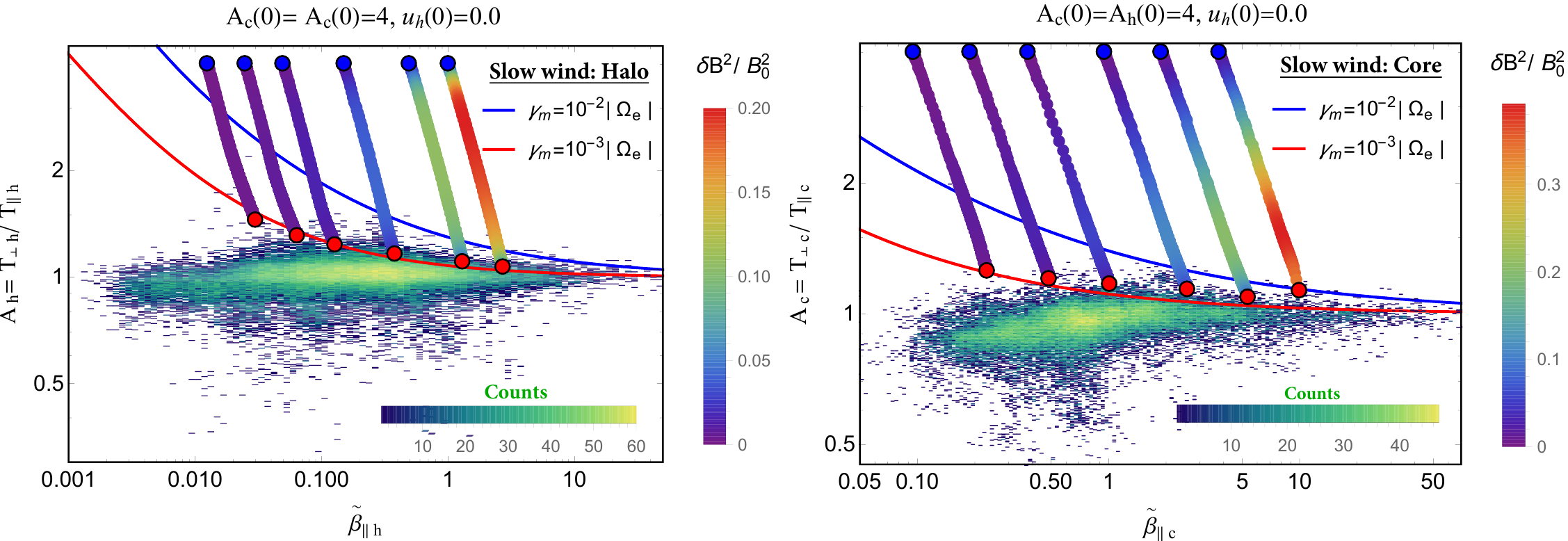}
\caption{Linear WI thresholds and QL dynamical paths of temperature anisotropy $A_h,c$ 
derived for $u_h=0.0$, are compared with the quasi-stable states of electron halo (left) and 
core (right) populations from slow winds ($V_{sw}<500$ km s$^{-1}$).} \label{Fig3}
   \end{figure*}
  \begin{figure*}[h!]
   \centering
\includegraphics[scale=0.75, trim={4.2cm 2.3cm 2.6cm 2.8cm}, clip]{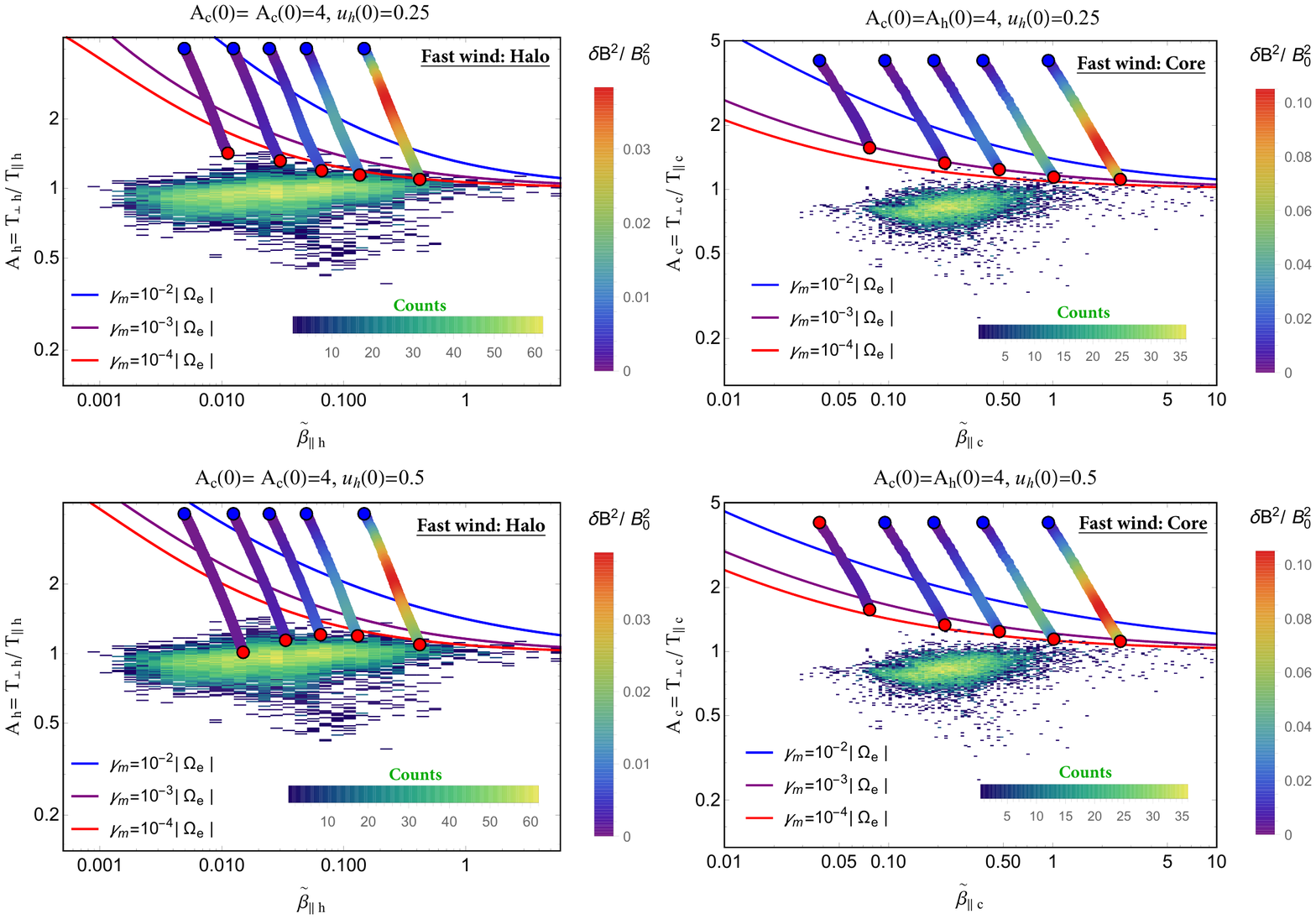}
\caption{Linear WI thresholds and QL dynamical paths of temperature anisotropy $A_h,c$ derived for
$u_h=0.25$ (top) and 0.5 (bottom), are compared with the quasi-stable states of electron halo (left) 
and core (right) populations from fast winds ($V_{sw}>500$ km s$^{-1}$).}   \label{Fig4}
   \end{figure*}

%%%%%%%%%%%%%%%%%%%%%%%%%%
\subsection{Constraints on the observed temperature anisotropies} \label{Sec:4.2}
%%%%%%%%%%%%%%%%%%%%%%%%%%
%
In order to emphasize the importance of the present study, in this section we perform a 
comparative analysis of the new anisotropy thresholds predicted by our QL approach 
for the cumulative whistler instability and temperature anisotropy of quasi-stable states, as 
reported by the observations of  solar wind electrons. The observational data set is selected 
from roughly 120 000 events detected in the ecliptic by three spacecrafts, \textit{Helios I},
\textit{ Cluster II}, and \textit{Ulysses} at different heliocentric distances from the Sun, 
i.e.\ $0.3-3.95$~AU. Figure~\ref{Fig3} and ~\ref{Fig4} display the observational data for the 
electron populations from slow (with $V_{SW}<500$ km s$^{-1}$) and fast winds (with $V_{SW}>500$ 
km s$^{-1}$), respectively. Distinction is also made between halo (left panels) and core (right
panels) electron populations. As already mentioned, these observational data are selected  
from the events associated with thermalized halo components (described by a $\kappa$-distribution 
with large enough $\kappa > 6$), for which the effects of the suprathermal populations on the QL 
saturation of WI and the relaxation of the anisotropic distribution are only minor 
\citep{Lazar2018b}. Occurrence rates showing the number of events in bins of temperature anisotropies 
$T_{\perp~a}/T_{\parallel~a}$ versus parallel plasma beta $\tilde{\beta}_{\parallel~a}\equiv n_a~
\beta_{\parallel~a}/n_e$ (subscript "$a=h$" for halo and "$a=c$" for core) are color-coded using a 
\textit{blue-green-yellow} color scheme. Unlike the results obtained by \cite{Adrian2016}, using a bulk model of the solar 
wind electrons, our calculations are based on a realistic dual model, which captures not only 
the effects of temperature anisotropy but also the influence of the core--halo relative
drift characteristic to fast winds.

The observational data are contrasted with the anisotropy thresholds of WI as resulted from the 
interplay of the halo and core electron populations for non-drifting velocities $u_h=~0$ in 
Figure~\ref{Fig3} and finite drifts $u_h\neq0$ in Figure~\ref{Fig4}. These thresholds are derived 
for different maximum growth rates $\gamma_m/ |\Omega_e|=~ 10^{-2}, 10^{-3}, 10^{-4}$, and are 
described with a general fitting law \citep{Lazar2014}
\begin{align}
A_{h,~c}=1+\frac{a}{\tilde{\beta}_{h,~c}^ {~b}}\left(1+\frac{c}{\tilde{\beta}_{h,~c}^ {~d}}\right). \label{e9}
\end{align}
Fitting parameters $a$, $b$, $c$ and $d$ are tabulated in Tables~\ref{t2}--\ref{t4}.

Figure~\ref{Fig3} enables a comparison of observational data, halo (left) and core (right) electron populations 
from slow winds ($V_{SW}<500$ km s$^{-1}$ and $u_a=~0$), with the WI thresholds and the QL dynamical paths 
obtained for different initial plasma betas, i.e., $\tilde{\beta}_{\parallel~h}(0)=~0.0125,~0.02,~0.05,~0.15,~0.5,~1$, and $\tilde{\beta}_{\parallel~c}(0)=~0.095,~ 0.19,~0.38,~0.95,~1.9,~3.8$, 
in a $(A_a, \tilde{\beta}_{\parallel~a})$ diagram. The initial positions, i.e.\ $A_a(0)-
\tilde{\beta}_{\parallel~a}(0)$, are marked with blue-filled circles, 
while after saturation the final positions, i.e.\ $A_a(\tau_m)-\tilde{\beta}_{\parallel~a}(\tau_m)$, are 
indicated by red-filled circles. The levels of the wave energy density $\delta B^2/ B_0^2$ are coded with 
\textit{rainbow} color scheme. We observe that the initial anisotropies are reduced in time toward the 
quasi-stable regimes and end up exactly to the WI threshold with a maximum growth rate $\gamma_m=10^{-3} 
|\Omega_e|$ (red line) predicted by the linear theory.
To be consistent with our dual model in Eq.~\eqref{e2} the electron data is more selective 
and more restrictive than data set used in \cite{Stverak2008}, but the new WI thresholds derived from a dual model 
show a clear trend to approach and shape the anisotropy limits for both the halo (left panel) and core 
(right panel) components. Moreover, final quasi-stable states resulted from dynamical paths of the halo (left) and 
core (right) anisotropies, can explain the observed more quasi-stable states of lower anisotropies.

Figure~\ref{Fig4} describes the effects of finite drift velocities $u_h(0)=~0.25$ (top) and $u_h(0)=0.5$ 
(bottom) on WI thresholds and the dynamical paths of temperature anisotropies, and contrast them 
with temperature anisotropies reported by the observations for the halo (left) and core (right) electrons.
We use the same $(A, \tilde{\beta}_{\parallel})$ diagrams assuming for the initial anisotropies $A_h(0)=
A_c(0)=4$. Dynamical paths are derived for different initial plasma betas, e.g., $\tilde{\beta}_{\parallel~h}(0)=~0.005,~0.015,
~0.025,~0.05,~0.15$. For the halo electrons the WI thresholds show the same clear trend to approach and shape 
the observed limits of temperature anisotropy. However, linear thresholds increase with increasing the 
halo drift velocity, i.e., from $u_h= 0.25$ to 0.5 [confirming the inhibiting effect of the halo drift 
velocity on the WI growth rates reported in \cite{Shaaban2018HFA}], that seems to reduce their relevance  
with respect to the observations for low $\tilde{\beta}_{h,\parallel}< 0.1$. As already discussed above, 
more relevant in this 
case are dynamical paths from QL theory, which allow to recover the quasi-stable states of lower 
anisotropies reported by the observations. These differences (including those determined by a 
variation of $u_h$) decrease with increasing $\beta_{\parallel~h}$. One possible explanation of this 
behavior can be given by the electrons with higher thermal velocities ($\beta_\parallel\propto v_\parallel^2$), 
which reduce the effectiveness of the halo drift velocity. 

The core electrons and the corresponding thresholds and dynamical paths are shown in Figure~\ref{Fig4}, 
right panels, using a $(A_c, \tilde{\beta}_{\parallel~c})$ diagram and the following initial plasma 
parameters $A_c(0)=A_h(0)=4.0$,  $u_h(0)=0.25$ (top),  $u_h(0)=0.5$ (bottom), and $\tilde{\beta}_{\parallel~c}(0)=~0.038,~0.095,~0.19,~0.38,~0.95$. The WI thresholds and dynamical paths show the same tendency 
to shape the limits of temperature anisotropy from observations, but a more satisfactory agreement is 
obtained only for a sufficiently high $\tilde{\beta}_{c,\parallel} > 0.5$, where the level of unstable
fluctuations is also markedly enhanced. The agreement between theory and observations is only qualitatively 
achieved in this case, suggesting that larger deviations from isotropy (which are not captured by our 
data of low time resolution) are actually constrained by the enhanced fluctuations resulting from a 
cumulative whistler instability. In the fast wind the core electrons may be more collisional retaining a 
higher thermalization from collisions, a feature that is missing in our collisionless plasma approach, but 
which may explain the near isotropic states from the observations \citep{Stverak2008, Yoon2016}. 
Instead, the halo electrons are more dilute and hotter, and therefore less affected by collisions but more 
susceptible to instabilities, which explain the good quantitative agreement obtained in left panels. In order to gain a reliable understanding on the interplay of electron core--halo 
relative drift and temperature anisotropy by isolating their effects on whistlers, in the present 
study we have analyzed only the events with a reduced influence of suprathermal electrons 
assuming both the core and halo Maxwellian distributed. Linear theory predicts a stimulation 
of kinetic instabilities in the presence of suprathermals, leading to higher growth rates and 
lower thresholds \citep{Vinas2015, Lazar2017a, Shaaban2018HF}. However, our present results
strongly suggest that future studies need to include these populations in extended QL and 
nonlinear approaches to provide a realistic picture of their implications.

%%%%%%%%%%%%%%%%%%%%%%%%%%%%%%%%
\section{Conclusions}\label{Sec:5}
%%%%%%%%%%%%%%%%%%%%%%%%%%%%%%%%
In the present manuscript we have characterized the long-run time evolution of whistler 
instability cumulatively driven by temperature anisotropy ($A_e = T_{e,\perp}/T_{e,\parallel}>1$) 
and counter-drifting electron populations, commonly encountered in the fast solar wind. 
The interplay of these two sources of free energy anisotropic temperature can markedly alter 
the QL evolution and saturation of the enhanced fluctuations, and implicitly the relaxation 
of the initial eVD. Studied here are the effects on the macroscopic plasma parameters such 
as plasma betas, temperature anisotropies, and drift velocities of both core and halo electrons. 

The relaxation of plasma parameters depends on the initial conditions, for which we considered 
two distinct cases. First we assumed an initially isotropic core $A_c (0) = T_{c,\perp}/T_{c,\parallel} = 
1$, motivated by the fact that the core electron population is a central component, much cooler 
and denser than the halo, and often showing lower deviations from isotropy. In the second case, 
both the core and halo were assumed having similar temperature anisotropies $A_c (0) \simeq A_h(0) > 1$. 
An initially isotropic core gains some energy from the RH-polarized electromagnetic fluctuations, 
and reaches a small temperature anisotropy in direction perpendicular to the magnetic field. 
The (counter-)drifting core-halo velocities $u_{c,h} \neq 0$ are another important factor that 
in general stimulates the relaxation of initial distributions, and implicitly the energy 
transfer through parallel heating and perpendicular cooling mechanisms. More exactly, it 
determines a faster relaxation, and increases the level reached by the fluctuating magnetic energy 
densities. However, the effects on temperature anisotropies are opposite, leading
to a more anisotropic core and a less anisotropic halo in the final states. For instance, 
in Figure 2, when both the core and halo populations are initially anisotropic, i.e., $A_c 
(0) \simeq A_h(0) > 1$, and $u_{h} = 0.5$, halo anisotropy after relaxation is distinctly 
much lower than that of the core. 
In all cases relaxation of the core and halo drifts is very modest, drifting velocities decreasing 
to values slightly lower than initial conditions. 

Section \ref{Sec:4.2} presents a comparative analysis of these new results predicted by a QL approach
for both  core and halo anisotropies, and the upper limits of the observed electron anisotropies 
in slow and fast solar winds. For non-drifting core and halo electron populations, i.e.\ $u_{c,h}=0$,
final quasi-stable states from dynamical paths of the temperature anisotropies align to the 
anisotropy limits of the core and halo anisotropies reported by the observations in the slow 
winds ($V_{SW}<500$ km~s$^{-1}$), and to anisotropy thresholds from linear theory \citep{Stverak2008}.
In the absence of relative drifts the instability thresholds predicted by linear theory coincide 
with the quasistable states resulted from a QL approach. In the fast winds, i.e.\ $V_{SW}>500$ 
km~s$^{-1}$, the relative drifts increase and cannot be neglected, i.e., $u_a\neq0$. Thresholds 
of WI increase moving away from the observed limits of temperature anisotropies 
with increasing the drift velocity, e.g., for \ $u_h=0.5$, and linear theory may show an agreement 
with the observations only for large $\tilde{\beta}_{\parallel~h}>0.1$ and very low growth rates 
$\gamma_m=10^{-4}|\Omega_e|$. The intensity of fluctuations so produced would be too low to account
for a constraining effect on particles. However, for the halo population, which is more susceptible
to kinetic instabilities, dynamical paths from our new QL approach reach more quasi-stable states 
of lower anisotropies, as reported by the observations in the fast winds. It seems also that 
this effect of whistler fluctuations on the temperature anisotropy of halo populations increases 
with increasing the drift velocity. The resolution of the invoked observations is below the one that may allow us to capture the unstable states, but as mentioned in the paper, these observations reveal the quasi-stable states expected to be found below the instability thresholds.

We conclude stating that anisotropy thresholds provided by linear theory for kinetic 
instabilities cumulatively driven by distinct sources of free energy may have a reduced 
relevance with respect to the observations, and therefore cannot explain the limits of 
the observed temperature anisotropy of electrons in the fast winds. Instead, we have 
shown that dynamical paths computed from an extended QL approach may provide a plausible
explanation for the quasistable states after the relaxation, and, implicitly, for the 
observations. Present analysis may be even more extended to include the contribution of 
collisions~\citep{Stverak2008, Yoon2016}, which may explain lower anisotropies of the core 
populations in the fast wind. Numerical simulations need to 
confirm the QL evolution and explain the nonlinear saturation of the instability (e.g., \cite{Moya2012, Seough2014, Moya2014, Lazar2018b}, but this will be the object 
of our future investigations.

%%%%%%%%%%%%%%%%   Table:2   %%%%%%%%%%%%%%%%%%
\begin{table}[h]
	\centering
	\caption{
	 Fitting parameters in Eq.(\ref{e9}) for $u_h=0.0$}
   \label{t2}
	\begin{tabular}{lcccccccc} % four columns, alignment for each	
		\hline
		&   \multicolumn{2}{c}{$\gamma_m=10^{-2} |\Omega_e|$} & \multicolumn{2}{c}{$\gamma_m=10^{-3} |\Omega_e|$} \\	
		 & Halo ($h$)  & Core ($c$)& Halo ($h$)& Core ($c$)\\
		\hline
		$a$ & 0.29 & 0.4& 0.08 & 0.11 \\
		$b$ & 0.45 & 0.45  &0.54 & 0.54\\
		$c$ & 0.0004& 0.0007 & $-10^{-5}$ & $-10^{-5}$\\
		$d$ & 1.0    & 1.0    & 1.0 & 1.0\\
		\hline
	\end{tabular}
\end{table}
%%%%%%%%%%%%%%%%%%%%%%%%%%%%%%%%%%%%%%%%%
%
%
%
%%%%%%%%%%%%%%%%   Table:3   %%%%%%%%%%%%%%%%%%
\begin{table}[h]
	\centering
	\caption{
	 Fitting parameters in Eq.(\ref{e9}) for $u_h=0.25$}
   \label{t3}
	\begin{tabular}{lcccccccc} % four columns, alignment for each	
		\hline
		&   \multicolumn{2}{c}{$\gamma_m=10^{-2} |\Omega_e|$} & \multicolumn{2}{c}{$\gamma_m=10^{-3} |\Omega_e|$} & \multicolumn{2}{c}{$\gamma_m=10^{-4} |\Omega_e|$}\\	
		 & Halo ($h$)  & Core ($c$)& Halo ($h$)  & Core ($c$)& Halo ($h$)  & Core ($c$)\\
		\hline
		$a$ & 0.28 & 0.39& 0.12 & 0.17  & 0.06 & 0.09 \\
		$b$ & 0.53 & 0.53  & 0.48 & 0.48  & 0.54 & 0.55\\
		$c$ & $-10^{-5}$& $-10^{-5}$ & $10^{-4}$ & $10^{-4}$&  $10^{-4}$ &  $-10^{-4}$\\
		$d$ & 1.0    & 1.0    & 1.0 & 1.0  & 1.0 & 1.0\\
		\hline
	\end{tabular}
\end{table}
%%%%%%%%%%%%%%%%%%%%%%%%%%%%%%%%%%%%%%%%%
%
%
%%%%%%%%%%%%%%%%   Table:4   %%%%%%%%%%%%%%%%%%
\begin{table}[h]
	\centering
	\caption{
	 Fitting parameters in Eq.(\ref{e9}) for $u_h=0.5$}
   \label{t4}
	\begin{tabular}{lcccccccc} % four columns, alignment for each	
		\hline
		&   \multicolumn{2}{c}{$\gamma_m=10^{-2} |\Omega_e|$} & \multicolumn{2}{c}{$\gamma_m=10^{-3} |\Omega_e|$} & \multicolumn{2}{c}{$\gamma_m=10^{-4} |\Omega_e|$}\\	
		 & Halo ($h$)  & Core ($c$)& Halo ($h$)  & Core ($c$)& Halo ($h$)  & Core ($c$)\\
		\hline
		$a$ & 0.41 & 0.53& 0.16 & 0.21  & 0.09 & 0.12 \\
		$b$ & 0.4 & 0.40  & 0.47 & 0.478  & 0.52 & 0.527\\
		$c$ & 0.00024& 0.005 & $-10^{-5}$ & $-10^{-5}$& $-10^{-5}$ &  $-10^{-4}$\\
		$d$ & 1.0    & 1.0    & 1.0 & 1.0  & 1.0 & 1.0\\
		\hline
	\end{tabular}
\end{table}
%%%%%%%%%%%%%%%%%%%%%%%%%%%%%%%%%%%%%%%%%

%%%%%%%%%%%%%%%%%%%%%%%%%%%%%%%%%%%%%%%%%%%%%%%%
\section*{Acknowledgements}

The authors acknowledge support from the Katholieke Universiteit Leuven, Ruhr-University Bochum, and 
Alexander von Humboldt Foundation. These results were obtained in the framework of the projects 
SCHL 201/35-1 (DFG-German Research Foundation), GOA/2015-014 (KU Leuven), G0A2316N (FWO-Vlaanderen), and C 90347 (ESA Prodex 9). S.M. Shaaban would like to acknowledge the support by a Postdoctoral Fellowship (Grant No. 12Z6218N) of the Research Foundation Flanders (FWO-Belgium). PHY acknowledges support from BK21 Plus project from NRF to Kyung Hee University. Part of his research was carried out during his visit to Katolieke Universiteit Leuven, Belgium. Stimulating discussion within the framework of ISSI project Kappa Distributions are gratefully appreciated.
Thanks are due to {\v S}.~{{\v S}tver{\'a}k} for providing the observational data. 
%%%%%%%%%%%%%%%%%%%%%%%%%%%%%%%%%%%%%%%%%%%%%%
\bibliographystyle{aa}

\bibliography{papers}

\begin{thebibliography}{36}
\expandafter\ifx\csname natexlab\endcsname\relax\def\natexlab#1{#1}\fi

\bibitem[{Adrian {et~al.}(2016)Adrian, Vi{\~{n}}as, Moya, \&
  Wendel}]{Adrian2016}
Adrian, M.~L., Vi{\~{n}}as, A.~F., Moya, P.~S., \& Wendel, D.~E. 2016, The
  Astrophysical Journal, 833, 49

\bibitem[{Ahmadi {et~al.}(2016)Ahmadi, Germaschewski, \& Raeder}]{Ahmadi2016}
Ahmadi, N., Germaschewski, K., \& Raeder, J. 2016, Journal of Geophysical
  Research: Space Physics, 121, 5350

\bibitem[{Bale {et~al.}(2013)Bale, Pulupa, Salem, Chen, \& Quataert}]{Bale2013}
Bale, S., Pulupa, M., Salem, C., Chen, C., \& Quataert, E. 2013, ApJL, 769, L22

\bibitem[{Fried \& Conte(1961)}]{Fried1961}
Fried, B. \& Conte, S. 1961, The Plasma Dispersion Function (New York: Academic
  Press)

\bibitem[{Gary(1985)}]{Gary1985}
Gary, S.~P. 1985, \jgr, 90, 10815

\bibitem[{Gary \& Karimabadi(2006)}]{Gary2006}
Gary, S.~P. \& Karimabadi, H. 2006, \jgr, 111, 1

\bibitem[{Lazar {et~al.}(2017b)Lazar, Pierrard, Shaaban, Fichtner, \&
  Poedts}]{Lazar2017b}
Lazar, M., Pierrard, V., Shaaban, S.~M., Fichtner, H., \& Poedts, S. 2017b,
  Astronomy {\&} Astrophysics, 602, A44

\bibitem[{Lazar {et~al.}(2014)Lazar, Poedts, \& Schlickeiser}]{Lazar2014}
Lazar, M., Poedts, S., \& Schlickeiser, R. 2014, \jgr, 119, 9395

\bibitem[{Lazar {et~al.}(2018a)Lazar, Shaaban, Fichtner, \&
  Poedts}]{Lazar2018a}
Lazar, M., Shaaban, S.~M., Fichtner, H., \& Poedts, S. 2018a, Phys. Plasmas, 25

\bibitem[{Lazar {et~al.}(2017a)Lazar, Shaaban, Poedts, \&
  {\v{S}}tver{\'{a}}k}]{Lazar2017a}
Lazar, M., Shaaban, S.~M., Poedts, S., \& {\v{S}}tver{\'{a}}k. 2017a, \mnras,
  464, 564

\bibitem[{Lazar {et~al.}(2018b)Lazar, Yoon, L{\'{o}}pez, \& Moya}]{Lazar2018b}
Lazar, M., Yoon, P.~H., L{\'{o}}pez, R.~A., \& Moya, P.~S. 2018b, \jgr, 123, 6

\bibitem[{L{\'{o}}pez {et~al.}(2019)L{\'{o}}pez, Lazar, Shaaban, Poedts, Yoon,
  Vi{\~{n}}as, \& Moya}]{Lopez2019}
L{\'{o}}pez, R.~A., Lazar, M., Shaaban, S.~M., {et~al.} 2019, 873, L20

\bibitem[{Maksimovic {et~al.}(2005)Maksimovic, Zouganelis, Chaufray, Issautier,
  Scime, Littleton, Marsch, McComas, Salem, Lin, {et~al.}}]{Maksimovic2005}
Maksimovic, M., Zouganelis, I., Chaufray, J.-Y., {et~al.} 2005, \jgr, 110,
  A09104

\bibitem[{Moya {et~al.}(2011)Moya, Mu{\~{n}}oz, Rogan, \& Valdivia}]{Moya2011}
Moya, P.~S., Mu{\~{n}}oz, V., Rogan, J., \& Valdivia, J.~A. 2011, Journal of
  Atmospheric and Solar-Terrestrial Physics, 73, 1390

\bibitem[{Moya {et~al.}(2014)Moya, Navarro, Vi{\~{n}}as, Mu{\~{n}}oz, \&
  Valdivia}]{Moya2014}
Moya, P.~S., Navarro, R., Vi{\~{n}}as, A.~F., Mu{\~{n}}oz, V., \& Valdivia,
  J.~A. 2014, \apj, 781, 76

\bibitem[{Moya {et~al.}(2012)Moya, Vi\~nas, Mu\~noz, \& Valdivia}]{Moya2012}
Moya, P.~S., Vi\~nas, A.~F., Mu\~noz, V., \& Valdivia, J.~A. 2012, Annales
  Geophysicae, 30, 1361

\bibitem[{{Pagel} {et~al.}(2007){Pagel}, {Gary}, {de Koning}, {Skoug}, \&
  {Steinberg}}]{Pagel2007}
{Pagel}, C., {Gary}, S.~P., {de Koning}, C.~A., {Skoug}, R.~M., \& {Steinberg},
  J.~T. 2007, \jgr, 112, A04103

\bibitem[{Pierrard {et~al.}(2016)Pierrard, Lazar, Poedts, {\v{S}}tver{\'{a}}k,
  Maksimovic, \& Tr{\'{a}}vn{\'{i}}{\v{c}}ek}]{Pierrard2016}
Pierrard, V., Lazar, M., Poedts, S., {et~al.} 2016, Solar Physics, 291, 2165

\bibitem[{Pulupa {et~al.}(2014)Pulupa, Bale, Salem, \& Horaites}]{Pulupa2014}
Pulupa, M.~P., Bale, S.~D., Salem, C., \& Horaites, K. 2014, \jgr, 119, 647

\bibitem[{Saeed {et~al.}(2017)Saeed, Sarfraz, Yoon, Lazar, \&
  Qureshi}]{Saeed2017a}
Saeed, S., Sarfraz, M., Yoon, P.~H., Lazar, M., \& Qureshi, M. N.~S. 2017,
  \mnras, 465, 1672

\bibitem[{Sarfraz {et~al.}(2016)Sarfraz, Saeed, Yoon, Abbas, \&
  Shah}]{Sarfraz2016}
Sarfraz, M., Saeed, S., Yoon, P.~H., Abbas, G., \& Shah, H.~A. 2016, \jgr, 121,
  9356

\bibitem[{Seough \& Yoon(2012)}]{Seough2012}
Seough, J. \& Yoon, P.~H. 2012, \jgr, 117, 1

\bibitem[{Seough {et~al.}(2014)Seough, Yoon, \& Hwang}]{Seough2014}
Seough, J., Yoon, P.~H., \& Hwang, J. 2014, Phys. Plasmas, 21, 062118

\bibitem[{Shaaban {et~al.}(2018a)Shaaban, Lazar, \& Poedts}]{Shaaban2018HF}
Shaaban, S.~M., Lazar, M., \& Poedts, S. 2018a, \mnras, 480, 310

\bibitem[{Shaaban {et~al.}(2018b)Shaaban, Lazar, Yoon, \&
  Poedts}]{Shaaban2018HFA}
Shaaban, S.~M., Lazar, M., Yoon, P.~H., \& Poedts, S. 2018b, Phys. Plasmas, 25,
  082105

\bibitem[{Shaaban {et~al.}(2019a)Shaaban, Lazar, Yoon, \&
  Poedts}]{Shaaban2019a}
Shaaban, S.~M., Lazar, M., Yoon, P.~H., \& Poedts, S. 2019a, \apj, 871, 237

\bibitem[{Shaaban {et~al.}(2019b)Shaaban, Lazar, Yoon, Poedts, \&
  L{\'{o}}pez}]{Shaaban2019b}
Shaaban, S.~M., Lazar, M., Yoon, P.~H., Poedts, S., \& L{\'{o}}pez, R.~A.
  2019b, arXiv, 10, 1

\bibitem[{Tong {et~al.}(2018)Tong, Bale, Salem, \& Pulupa}]{Tong2018}
Tong, Y., Bale, S.~D., Salem, C., \& Pulupa, M. 2018, arXiv, 1

\bibitem[{Tong {et~al.}(2019)Tong, Vasko, Pulupa, Mozer, Bale, Artemyev, \&
  Krasnoselskikh}]{Tong2019}
Tong, Y., Vasko, I.~Y., Pulupa, M., {et~al.} 2019, ApJL, 870, L6

\bibitem[{{{\v S}tver{\'a}k} {et~al.}(2008){{\v S}tver{\'a}k},
  {Tr{\'a}vn{\'{\i}}{\v c}ek}, {Maksimovic}, {Marsch}, {Fazakerley}, \&
  {Scime}}]{Stverak2008}
{{\v S}tver{\'a}k}, {\v S}., {Tr{\'a}vn{\'{\i}}{\v c}ek}, P., {Maksimovic}, M.,
  {et~al.} 2008, \jgr, 113, A03103

\bibitem[{Vi{\~n}as {et~al.}(2010)Vi{\~n}as, Gurgiolo, Nieves-Chinchilla, Gary,
  \& Goldstein}]{Vinas2010}
Vi{\~n}as, A., Gurgiolo, C., Nieves-Chinchilla, T., Gary, S.~P., \& Goldstein,
  M.~L. 2010, AIP Conference Proceedings, 1216, 265

\bibitem[{Vi{\~{n}}as {et~al.}(2015)Vi{\~{n}}as, Moya, Navarro, Valdivia,
  Araneda, \& Muñoz}]{Vinas2015}
Vi{\~{n}}as, A.~F., Moya, P.~S., Navarro, R.~E., {et~al.} 2015, Journal of
  Geophysical Research: Space Physics, 120, 3307

\bibitem[{{Vocks} {et~al.}(2005){Vocks}, {Salem}, {Lin}, \& {Mann}}]{Vocks2005}
{Vocks}, C., {Salem}, C., {Lin}, R.~P., \& {Mann}, G. 2005, \apj, 627, 540

\bibitem[{Yoon(1992)}]{Yoon1992}
Yoon, P.~H. 1992, Physics of Fluids B: Plasma Physics, 4, 3627

\bibitem[{Yoon(2016)}]{Yoon2016}
Yoon, P.~H. 2016, Phys. Plasmas, 23, 072114

\bibitem[{Yoon {et~al.}(2012)Yoon, Seough, Kim, \& Lee}]{Yoon2012}
Yoon, P.~H., Seough, J.~J., Kim, K.~H., \& Lee, D.~H. 2012, J. Plasma Phys.,
  78, 47

\end{thebibliography}

\end{document}